\documentclass[showpacs, preprint,amsmath,amssymb,aps]{revtex4}
\usepackage[T1]{fontenc}
\usepackage[utf8]{inputenc}
\usepackage{graphicx}
\usepackage{dcolumn}
\usepackage{bm}
\usepackage[version=3]{mhchem}
\usepackage{tabularx}
\usepackage{booktabs}
\usepackage{url}
\usepackage{hyperref}
\usepackage{color}

\begin{document}


\title{Impurity scattering and size quantization effects in a single graphene nanoflake}

\author{Julia Tesch$^{1}$}
\author{Philipp Leicht$^{1}$}
\author{Felix Blumenschein$^{1}$}
\author{Luca Gragnaniello$^{1}$}
\author{Anders Bergvall$^{2}$}
\author{Tomas L\"ofwander$^{2}$}
\author{Mikhail Fonin$^{1}$\footnote{Email address: mikhail.fonin@uni-konstanz.de}}

\affiliation{
$^{1}$~\mbox{Fachbereich Physik, Universit\"at Konstanz, 78457 Konstanz, Germany}\\
}

\affiliation{
$^{2}$~\mbox{Department of Microtechnology and Nanoscience}\\\mbox{MC2, Chalmers University of Technology, SE-412 96 G\"oteborg, Sweden}\\
}

\date{\today}

\begin{abstract}

By using Fourier-transform scanning tunneling spectroscopy we measure the interference patterns produced by the impurity scattering of confined Dirac quasiparticles in epitaxial graphene nanoflakes. Upon comparison of the experimental results with tight-binding calculations of realistic model flakes, we show that the characteristic features observed in the Fourier-transformed local density of states are related to scattering between different transverse modes (sub-bands) of a graphene nanoflake and allow direct insight into the electronic spectrum of graphene. We also observe a strong reduction of quasiparticle lifetime which is attributed to the interaction with the underlying substrate. In addition, we show that the distribution of the onsite energies at flower defects leads to an effectively broken pseudospin selection rule, where intravalley back-scattering is allowed. 
\end{abstract}

\pacs{73.20.At, 73.22.Pr, 68.37.Ef, 72.10.Fk}

\maketitle

Following the exfoliation of graphite monolayers in 2004~\cite{Novoselov:2004}, graphene has attracted considerable interest as a prospective material for electronic~\cite{Geim:2007,CastroNeto:2009} and spintronic \cite{Tombros:2007,Avsar:2011} applications. Upon fabrication of graphene nanoribbons~\cite{Tapaszto:2008,Wang:2010,Ruffieux:2010,Tao:2011,Lit:2013,Baringhaus:2014} or nanoislands~\cite{Morgenstern:2016,Tesch:2016} with different edge terminations, new properties can be introduced  such as tunable bandgaps~\cite{Nakada:1996,Son:2006a,Yang:2007} or edge-induced magnetism~\cite{Son:2006b,Yazyev:2010}. With regard to the electronic transport through graphene nanoribbons, the effect of impurity scattering and edge disorder becomes an important issue. A powerful tool to examine the quasiparticle interference (QPI) effects in graphene due to scattering at defects and edges is scanning tunneling microscopy and spectroscopy~\cite{Rutter:2007,Mallet:2007,Brihuega:2008,Xue:2012}. The observed QPI is directly related to modulations in the local density of states (LDOS)~\cite{Friedel:1958,Simon:2007} and provides access to the present scattering vectors and thus to the electronic structure of graphene~\cite{Bena:2008,Pereg:2008,Mallet:2012,Simon:2011}. Recently, the influence of local scattering centers on the local density of states in graphene nanoribbons has been studied theoretically~\cite{Bergvall:2013}. The interplay between single impurity scattering and size quantization was shown to generate characteristic spectral features in the Fourier Transform (FT) LDOS that can be related to the transverse modes of the nanoribbon. 

Here we present a comprehensive study of size quantization in epitaxial graphene nanoflakes (GNFs) on Ag(111) upon analysis of QPI by STM and tight-binding simulations of realistic model flakes. We indeed find the characteristic features in the FT-LDOS related to scattering between different transverse modes of a GNF as predicted by theory. Detailed analysis of the scattering features allows to gain a profound insight into the behavior of charge carriers in graphene flakes, including discrete electronic spectrum, quasiparticle lifetimes as well as effects of pseudospin.

Graphene nanoflakes were initially grown on Ir(111) and decoupled by noble metal intercalation as described elsewhere~\cite{Leicht:2014,Tesch:2016}. STM and STS measurements were carried out in an \textit{Omicron} cryogenic STM setup in ultra-high vacuum at $T$=5--10\,K. Differential conductance (d$I$/d$V$) maps were obtained using a standard lock-in technique with modulation voltages $V_\text{mod}$ = 3\,mV(rms) and at frequencies $f_{\text{mod}}$ = 600--800\,Hz. Tight-binding calculations were performed using an atomistic recursive Green's function formalism including the effects of trigonal warping in order to account for the relatively large doping level of graphene on Ag(111). AB sublattice symmetry breaking was introduced via rotational grain boundary defects modeled as a local $\sigma_z$ (mass) potential in addition to a scalar potential. Tight-binding parameters were inferred from DFT simulations of comparable topological defects. The limited quasiparticle lifetime in modeling realistic flakes was achieved by an increase of the imaginary part of the energy above the energy level separation resulting in the loss of transverse quantization within the flake. 

\begin{figure}
\includegraphics[width=8.7cm]{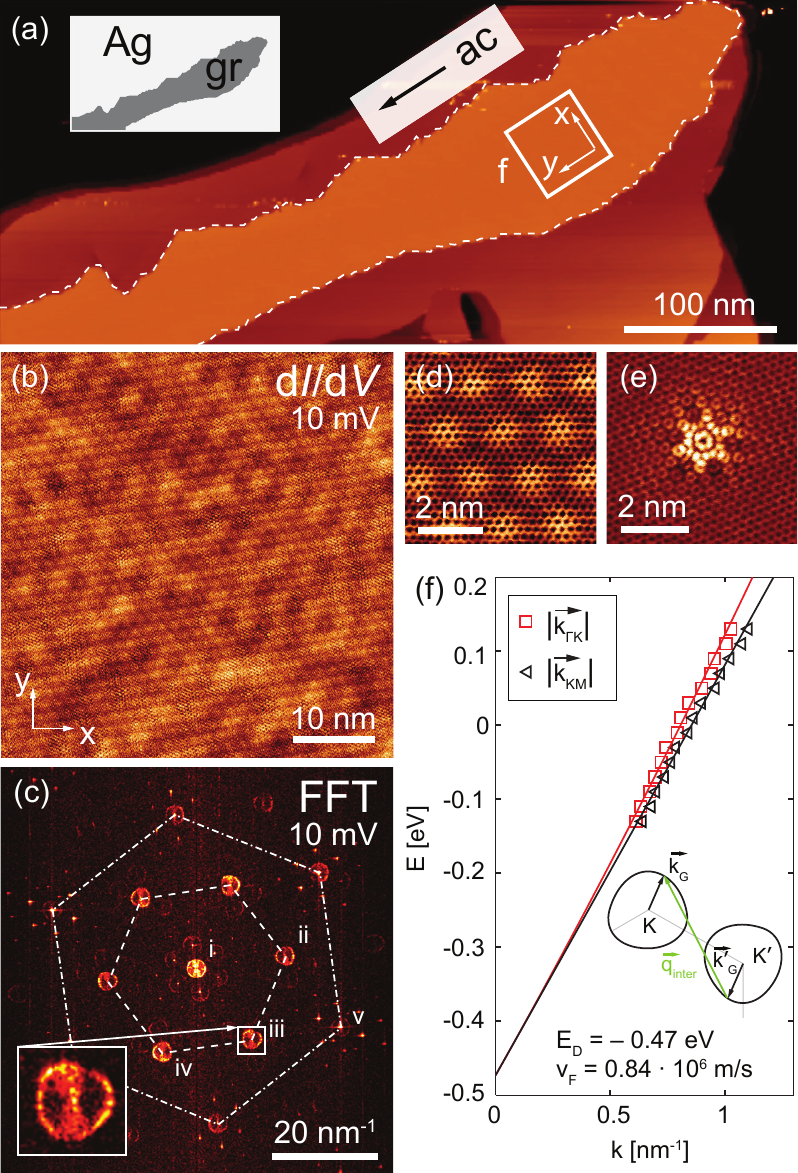}
\caption{\label{STM_and_STS_on_the_Long_Free} (a) STM image of a GNF on Ag(111) ($V$ = 0.1\,V; $I$ = 300\,pA). (b) d$I$/d$V$ map of the area marked in (a) recorded at $V = 10$\,mV. (c) Corresponding FFT of the conductance map displaying intravalley (i) and intervalley (ii-iv) scattering features as well as reciprocal lattice and Moir\'e spots. Inset: Close up of the intervalley scattering feature (iii). (d) Atomically resolved image showing Moir\'e and atomic lattice ($V$ = 0.1\,V; $I$ = 1.5\,nA). (e) STM image of a flower defect ($V$ = -0.2\,V; $I$ = 800\,pA). (f) Dispersion relation of the graphene flake shown in (a) with respect to the \textit{K} point, as derived from d$I$/d$V$ maps. Inset: Schematic of the intervalley scattering process.}
\end{figure}

Figure~\ref{STM_and_STS_on_the_Long_Free} (a) shows  a typical elongated graphene nanoflake sitting on top of an at least 7\,nm high epitaxial Ag(111) island on Ir(111). Prominent LDOS modulations, modifying the honeycomb appearance of graphene are observed in d$I$/d$V$ mappings [Fig.~\ref{STM_and_STS_on_the_Long_Free} (b)] leading to pronounced features in the fast Fourier transform (FFT) image [Fig.~\ref{STM_and_STS_on_the_Long_Free}(c)]. The LDOS patterns at the edges are identical to those previously reported for GNFs on Au(111)~\cite{Leicht:2014}, suggesting single H-terminated graphene edges. The analysis of the atomic contrast and Moir\'e structure [Fig.~\ref{STM_and_STS_on_the_Long_Free}(d)] suggests the $R$0 adsorption configuration of the GNF, meaning that the metal $<11\bar{2}>$ direction is parallel to the graphene $<1\bar{1}00>$ direction. Atomic reconstruction reveals an alignment of the long flake axis roughly parallel to the armchair (ac) direction and a predominant termination with zigzag (zz) edges for both sides of the flake. Besides the edges, point defects within the flake interior also give rise to strong LDOS modulations. The most abundant defect type observed in the experiments are flower defects~\cite{Cockayne:2011} with the atomic appearance shown in Fig.~\ref{STM_and_STS_on_the_Long_Free} (e).

Figure~\ref{STM_and_STS_on_the_Long_Free} (b) displays a d$I$/d$V$ map on the GNF recorded at 10\,mV. As the electronic band minimum of the Ag(111) surface state is shifted upwards to about 200\,meV due to the presence of graphene~\cite{Leicht:2014,Jolie:2015} as well as due to strain effects in the Ag thin film~\cite{Neuhold:1997,Tomanic:2012}, we can unambiguously assign the observed LDOS modulations to graphene standing waves superposed by the Moir\'e superstructure. The corresponding FFT of the $\text{d}I/\text{d}V$ map [Fig.~\ref{STM_and_STS_on_the_Long_Free} (c)] reveals intervalley and intravalley scattering features characteristic for graphene~\cite{Mallet:2012}. The observed intravalley feature stems from the elastic scattering between the states on a constant energy contour (CEC) within a single Dirac cone, whereas the intervalley features correspond to the scattering vectors $\vec{q}_{\text{inter}}$ connecting the states of the two neighboring cones~\cite{Mallet:2012} as displayed in Fig.~\ref{STM_and_STS_on_the_Long_Free} (f). Due to the charge transfer leading to the shift of the Dirac point to -470\,mV [Fig.~\ref{STM_and_STS_on_the_Long_Free} (f)], scattering features for graphene on Ag(111) already show the onset of the trigonal warping. The appearance of the intervalley features deviates substantially from the ones reported for a perfect infinite graphene sheet, where an intensity modulated ring-like contour is expected~\cite{Rutter:2007,Brihuega:2008,Mallet:2012}. Instead, our measurements show a ring-like feature with considerable intensity inside the ring, which is located on a line through the center of the scattering ring [inset in Fig.~\ref{STM_and_STS_on_the_Long_Free} (c)]. The alignment of this feature reflects the direction of the long axis of the GNF in real space. In contrast to that no inner structure within the intervalley features is observed in measurements on extended graphene on Ag(111) underlining the confinement nature of the observed features.

\begin{figure}
\includegraphics[width=8.7cm]{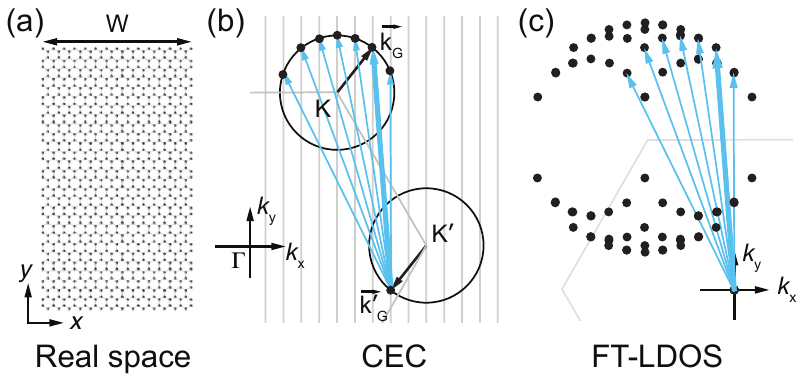}
\caption{\label{Model} (a) Real space geometry of a metallic AGNR of width W. (b) Schematic representation of the constant energy contours of an AGNR with possible scattering processes (intervalley) indicated by arrows. Due to the transverse confinement $k_x = n\pi/W - K_{dx}$ is quantized giving rise to a number of scattering vectors originating at $\vec{k}'_G$ (blue arrows) as compared to an infinite graphene sheet, where only points with antiparallel vectors $\vec{k}_G$ and $\vec{k}'_G$ can be connected (thick blue arrow). (c) Resulting FT-LDOS of a metallic AGNR corresponding to the transitions depicted in (b).}
\end{figure}

In order to understand the effect of quantum confinement on the electron scattering, we first discuss the simplified case of an infinitely long armchair graphene nanoribbon (AGNR) of width $W$ [Fig.~\ref{Model} (a)]. With $x$ corresponding to the confined direction, the transverse wave vector $k_x$ is quantized and can be written as $k_x \rightarrow k_{dn} = n\pi/W-K_{dx}$, where $n$ is an integer and $d$ enumerates valley pairs in the first Brillouin zone~\cite{Bergvall:2013}. More precisely, the values of $k_x$ involved in a scattering event at a point defect in the ribbon are located at intersections of lines in the $k_y$ direction with the CEC of graphene at positions $k_{dn}$ [Fig.~\ref{Model} (b)]. In a ribbon, scattering vectors are observed connecting two points on the circular CEC with initial values $k_{dn} = n\pi/W-K_{dx}$ and final values $k_{d'm}=m\pi/W-K_{d'x}$ (specific selection rules must also be obeyed)~\cite{Bergvall:2013}. Thus, scattering intensity is observed only at a finite number of $\vec{q}$ points, which lie inside the circle of radius 2$\sqrt{2mE/\hbar^2}$ and constitute a scattering feature as shown in Fig.~\ref{Model} (c). When comparing these characteristic patterns with the experimentally observed scattering features of the GNF along the armchair direction [inset in Fig.~\ref{STM_and_STS_on_the_Long_Free} (c)], there is already a striking resemblance for both contour and interior.

\begin{figure}
\includegraphics[width=8.7cm]{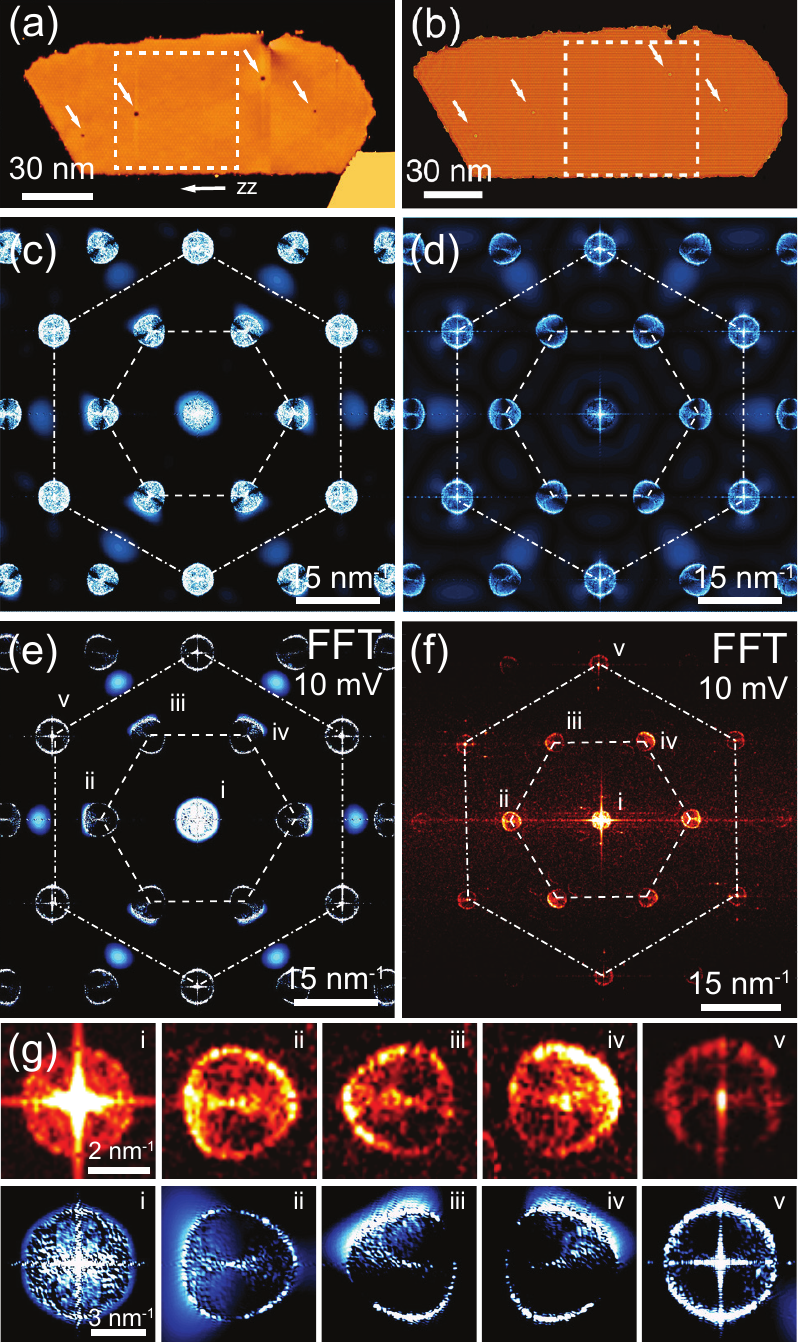}
\caption{\label{STM_and_STS_on_RectFlake} (a) STM image of a compact GNF ($V$ = 0.03\,V; $I$ = 300\,pA) and long axis along the zigzag direction. (b)  Atomic model of a quasi free-standing flake as used in the tight binding calculations. (c) FT-LDOS of the region of the model flake marked in (b) assuming infinite quasiparticle lifetime and no symmetry breaking at grain boundary sites. The intervalley scattering interior arising from confinement is directed towards the centre of the FFT. (d) FT-LDOS including limited quasiparticle lifetime, thus yielding intervalley features aligned with the flake's long axis inside the scattering contour. (e) FT-LDOS including limited quasiparticle lifetime as well as AB sublattice symmetry breaking due to the implementation of a mass term at the rotational grain boundary sites. (f) FFT of the experimental d$I$/d$V$ map of the flake region marked in (a) exhibiting intravalley scattering (i), intervalley scattering (ii-iv) and higher order features (v). (g) Magnification of the scattering features visible in the experimental and theoretical FT-LDOS.}
\end{figure}

Going back to scattering in realistic graphene structures, we extend the analysis to a direct comparison of the experimental data from a compact GNF with the results of a theoretical treatment of this particular flake geometry. Fig.~\ref{STM_and_STS_on_RectFlake} (a) shows a $R$0 GNF on Ag(111) with a lateral size of 70$\times$170\,nm$^2$. The long edges of the experimentally measured flake are virtually parallel to the zigzag direction, resembling the shape of a zigzag graphene nanoribbon with rough edges. The size of the flake allows for a reasonable atomistic tight-binding modeling of graphene $\pi$-electrons, utilizing a numerical recursive Green's function approach.  In this approach, we assume a free-standing graphene flake with about $3.4\cdot 10^5$ atoms [Fig.~\ref{STM_and_STS_on_RectFlake} (b)] reproducing the overall size and shape of the experimental GNF, with the atomic structure precisely adjusted to the experimentally observed one including flower defects and edges. Since the edges are hydrogen passivated, each edge carbon atom is simply modeled by a single $\pi$-orbital as in the flake interior. The results of this approach are in very good agreement with previous experimental observations and theoretical calculations for the edges~\cite{Leicht:2014}.

The tight-binding model used is defined by the Hamiltonian
\begin{equation}
H  = \sum_{i}\epsilon_ic_i^{\dagger}c_i +  \sum_{ij} t_{ij}c_i^{\dagger}c_j,
\end{equation}
where $c_i^{\dagger}$ and $c_i$ are creation and annihilation operators for site $i$, $\epsilon_i$ is the onsite energy of site $i$, and $t_{ij}$ is the hopping amplitude between sites $j$ and $i$. We set the tight-binding hopping parameters according to the simple formula for $\pi$ orbital overlap~\cite{LofwanderPRB2014}
$t_{ij} = t(r) = -t \exp[-\lambda(r-a_{cc})]$,
where $r$ is the distance in the plane between carbon atoms $i$ and $j$, $a_{cc}=1.42$~\AA~is the graphene carbon-carbon distance, and $\lambda\approx 3/a_{cc}$. The nearest-neighbor hopping integral $-t$ is related to the Dirac electron Fermi velocity $\hbar v_F=3a_{cc}t/2$. From the measured Fermi velocity [Fig.~\ref{STM_and_STS_on_the_Long_Free} (f)], we get $t\approx 2.6$~eV. The above formula is applied for inter-atomic distances $r$ less than a cut-off $R_c\approx 1.8 a_{cc}$, in order to take into account next-nearest neighbor hopping as well as reasonable hopping parameters in the pentagons and heptagons in the flower defects. We set onsite energies $\epsilon_i$ to zero throughout the flake, except in the flower defects where they vary between $-0.6t$ and $+1.2t$. This serves to model the charge transfer between pentagons and heptagons in the defect. The onsite energies are unknown for the flower defect and should in principle be computed from density functional theory. Here the parameters have been taken in analogy to the Stone-Wales defect, for which this has been done thoroughly by Amara \emph{et al.}~\cite{AmaraPRB2007}. For the FT-LDOS, the exact values of these parameters actually do not matter, as long as they are non-zero. For zero onsite energies across the flower defect, the scattering is too weak to match the experiment. We note that the distribution of onsite energies in the flower defect leads to an effectively broken pseudospin selection rule, where intravalley back-scattering is allowed. This leads to the restoration of the intravalley scattering ring in the FT-LDOS (feature i in Fig.~\ref{STM_and_STS_on_RectFlake} (e), which can also be observed experimentally in both armchair and zigzag GNFs (see Fig.~\ref{STM_and_STS_on_the_Long_Free} and \ref{STM_and_STS_on_RectFlake} for a comparison). The presence of flower defects in the flake also yields additional intensities besides scattering and atomic features in the FFT related to the defect geometry, which are found for both the experimental and theoretical flake, however, they are more pronouncedly visible in the tight-binding FFT due to the absence of measurement noise. Taking into account next-nearest neighbor hopping $t(r=\sqrt{3}a_{cc})=-t'$, we have in the above model $E_D=3t'\approx 0.33t$. For the simulations we use $E_F=0.61t$, which is more electron-doped than observed experimentally. This serves to enlarge the scattering features, see Fig.~\ref{STM_and_STS_on_RectFlake}. It should be noted that this puts us further into the trigonal warping regime due to the substantial shift of E$_D$ with respect to E$_F$.

The LDOS is computed via the imaginary part of the retarded Green's function 
$G=\left(E_F+i\eta - H\right)^{-1}$.
Broadening of energy levels can lead to loss of visibility of the scattering features (c.f. Fig.~\ref{Model}) associated with size quantization. Broadening can be due to electron-electron interaction, electron-phonon interaction or, for instance, weak coupling of the graphene $\pi$-electron system to the underlying metallic substrate. Here we introduce it phenomenologically as an imaginary part of the energy $E\rightarrow E+i\eta$ when computing the Green's function for electron propagation in the flake. For example, for a rectangular flake of dimension 70\,nm $\times$ 170\,nm, with the zigzag direction along the long axis, we have different estimates for energy level separations in the two directions. We get $\Delta E_{ac} \approx 10$\,meV and $\Delta E_{zz} \approx 25$\,meV. Thus, at an imaginary part $\eta\sim \Delta E_{ac}$, we will no longer see quantization along the long axis, only along the short axis. Indeed, the simulated FT-LDOS images show complicated flake levels for small $\eta$, while these are broadened in favor of clean zigzag ribbon like levels for the chosen $\eta=3\cdot 10^{-3}t\approx 8$\,meV, see Fig.~\ref{STM_and_STS_on_RectFlake} (e). This translates into a quasiparticle lifetime of about 80\,fs and, taking into account the experimentally obtained Fermi velocity, a mean free path of about 70\,nm, which corresponds to the short axis length of the flake. In comparison to extended graphene sheets, the obtained quasiparticle lifetime is approximately one order of magnitude smaller\cite{Andrei:2009}. 

Upon examination of the intervalley scattering features in the experimentally obtained FFT [Fig.~\ref{STM_and_STS_on_RectFlake} (f)], in addition to graphene bulk features (circles), a rich inner structure due to transverse confinement is visible in the FT-LDOS. The calculated FT-LDOS [Fig.~\ref{STM_and_STS_on_RectFlake} (e)] displays very pronounced trigonally warped intervalley scattering contours enclosing additional features aligned with the flake's long axis. These fine structures are very narrow at the center of the ring, while fanning out towards the rim. This is in perfect agreement with the experimentally obtained data as can be readily seen from the juxtaposition of the magnified experimental and theoretical scattering features compiled in Fig.~\ref{STM_and_STS_on_RectFlake} (g). Also the intravalley scattering ring observed in the measured FT-LDOS is restored in the tight-binding calculation upon the inclusion of AB sublattice symmetry breaking at the flower defects. Thus a correct description of the scattering features is only possible when taking into account the size and exact atomic structure of the nanoflake along with the symmetry-breaking at point defects and substantial energy level broadening.

In summary, clear footprints of confinement can be observed in quasiparticle scattering of epitaxially grown graphene nanoflakes. We have demonstrated that the experimentally observed scattering features are in very good agreement with tight-binding calculations of well-defined graphene nanoribbons, which besides conventional intra- and intervalley scattering display quasiparticle wave vectors arising from additional scattering channels between the ribbons' transverse modes.  However, certain interactions and imperfections of the crystal lattice in realistic flakes have to be accounted for. Modeling flakes with grain boundary defects allows us to distinguish the effects of rough edges, defects and quasiparticle lifetime. While we have shown that the introduced grain boundary defect breaks the AB sublattice symmetry thus restoring the intravalley scattering contour at $\vec{q}=0$, it can not reproduce the alignment of the scattering feature interior with the long axis of the calculated flakes. This was achieved upon implementation of a finite quasiparticle lifetime in analogy to the scattering processes in infinite graphene nanoribbons. Due to the limited lifetime, confinement is hence only preserved along the short axis visible in the FT-LDOS as an altered intensity distribution strongest at the centre of the ring contours and aligned with the long axis.\\

The authors gratefully acknowledge financial support by the Deutsche Forschungsgemeinschaft within the Priority Program (SPP) 1459,
the Swedish foundation for strategic research (SSF), and Knut and Alice Wallenberg foundation (KAW).

\end{document}